# Deep Open Snake Tracker for Vessel Tracing

Li Chen[1][0000-0003-0233-4576], Wenjin Liu[1], Niranjan Balu[1],

Mahmud Mossa-Basha[1], Thomas S. Hatsukami[1], Jenq-Neng Hwang[1], Chun Yuan[1]

[1] University of Washington, Seattle WA 98109, USA
`cluw@uw.edu`

**Abstract.** Vessel tracing by modeling vascular structures in 3D medical images with centerlines and radii can provide useful information for vascular health. Existing algorithms have been developed but there are certain persistent problems such as incomplete or inaccurate vessel tracing, especially in complicated vascular beds like the intracranial arteries. We propose here a deep learning based open curve active contour model (DOST) to trace vessels in 3D images. Initial curves were proposed from a centerline segmentation neural network. Then data-driven machine knowledge was used to predict the stretching direction and vessel radius of the initial curve, while the active contour model (as human knowledge) maintained smoothness and intensity fitness of curves. Finally, considering the non-loop topology of most vasculatures, individually traced vessels were connected into a tree topology by applying a minimum spanning tree algorithm on a global connection graph. We evaluated DOST on a Time-of-Flight (TOF) MRA intracranial artery dataset and demonstrated its superior performance over existing segmentation-based and tracking-based vessel tracing methods. In addition, DOST showed strong adaptability on different imaging modalities (CTA, MR T1 SPACE) and vascular beds (coronary arteries).

## 1    Introduction

Cardiovascular disease is one of the leading causes of death and disability worldwide, primarily via coronary artery disease and stroke [1]. Medical imaging such as computed tomography angiography (CTA) and magnetic resonance angiography (MRA) allows visualization of vasculatures. Through 3D vascular map construction, topological and morphometric information can be quantified, providing vascular information for clinical diagnosis and vascular health assessment, including the presence of stenosis, occlusion, collateral arteries, and distribution [2-6].

A critical and challenging step for vascular map construction is accurate and automated vessel tracing, which is converting original vascular images into a vascular network with topological representation of branches, including centerlines and radii. In general, two main approaches exist for automated vessel tracing. **1) Segmentation-based approaches.** Voxels belonging to the vascular region are segmented, then the vessel skeletonization method identifies the centerline through iterative thinning. Finally, radii along the centerlines are estimated. Methods belonging to this approach mainly differ in the segmentation algorithms, which are comprehensively discussed in review papers [7–9]. Recently, convolutional neural network (CNN) based vascular segmentation has become the predominant method, such as using the patch origin encoded Y-Net [10] and distance transformed segmentation [11]. For segmentation-based approaches, there is no guarantee for the smoothness and continuity of vessels after skeletonization. Moreover, two nearby vessels close to each other might be traced as one large vessel. **2) Tracking-based approaches.** Initial seeds are identified as the starting points from the vascular images, then the artery centerline and radius are directly identified from seed points through iteratively stretching both ends of the trace during tracking. It is critical to predict the correct direction for stretching, which can either be from a human designed vesselness enhancement filter (for example, the first principal direction of the Jacobian matrix of images [12]), or from a neural network used in CNN Tracker and DCAT [13,14]. The tracking-based approach is sensitive to initial seed placement. Improper seed selection is likely to cause tracing leakage into the background or result in an incomplete vascular tree. Traces might also be rough or zigzagged due to a lack of smoothness constraints.

In this study, we proposed a deep open active contour (snake) tracker (DOST), a hybrid method taking advantage of both segmentation and tracking approaches for robust vessel tracking. More importantly, DOST, by merging deep learning predictions into the traditional open snake algorithm, combines human knowledge (a snake algorithm to ensure the smoothness and intensity fitness of traces) and data-driven machine knowledge (deep learning to ensure reliable and robust vascular tracing) to construct a topologically correct vascular tree.

The main innovations and contributions of our work include: 1) DOST, proposed to solve automated vessel tracing; 2) by benchmarking with an intracranial MRA dataset, we achieved the state-of-the-art performance on vessel tracing; 3) DOST is applicable to multiple vascular beds and imaging modalities; 4) new evaluation metrics were used to evaluate multi-vessel connection accuracy of complicated vascular structures, a supplement to existing overlap-based metrics.

## 2 Methods

### 2.1 Deep open curve snake

Active contour model, also called snake [15], was first introduced as an algorithm for contour delineation on noisy 2D images. Initial contours were provided by users, then contours are refined through iterative minimization of the energy function, considering the external energy for fitting image intensity and the internal energy conserving contour smoothness. Wang et al [12,16] extended snake to open curve snake (OCS) for 3D tubular neural fibers tracing by optimizing on open traces (initialized from seed points) instead of closed contours. However, three problems limit the use of OCS on more complicated vessel tracing: 1) seed initialization and initial stretching direction are sensitive to noise; 2) stretching directions are decided purely by the local image gradients, which is sensitive to noise and not applicable to vessel bifurcation/branches; 3) each trace is independently identified without global constraints. Connection mistakes such as loops might appear in the vascular tree (**Fig. 1** (a)).

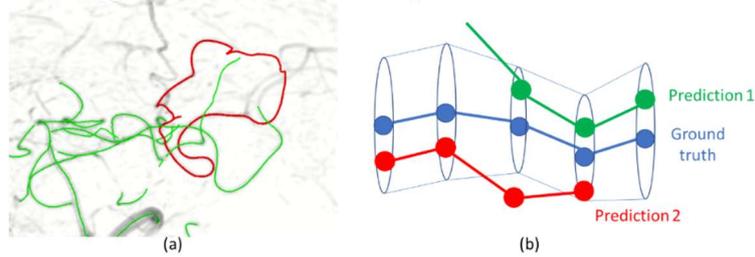

**Fig. 1.** (a) An example of loop (red) due to errors in tracing intracranial arteries using the OCS method. (b) An example when overlap measures are high but traced by multiple traces.

Deep open snake tracker (DOST) is designed to improve OCS with deep learning and global tree structure constraints. DOST includes key steps: curve proposal from centerline segmentation, deep snake tracing, and global tree construction (**Fig. 2**).

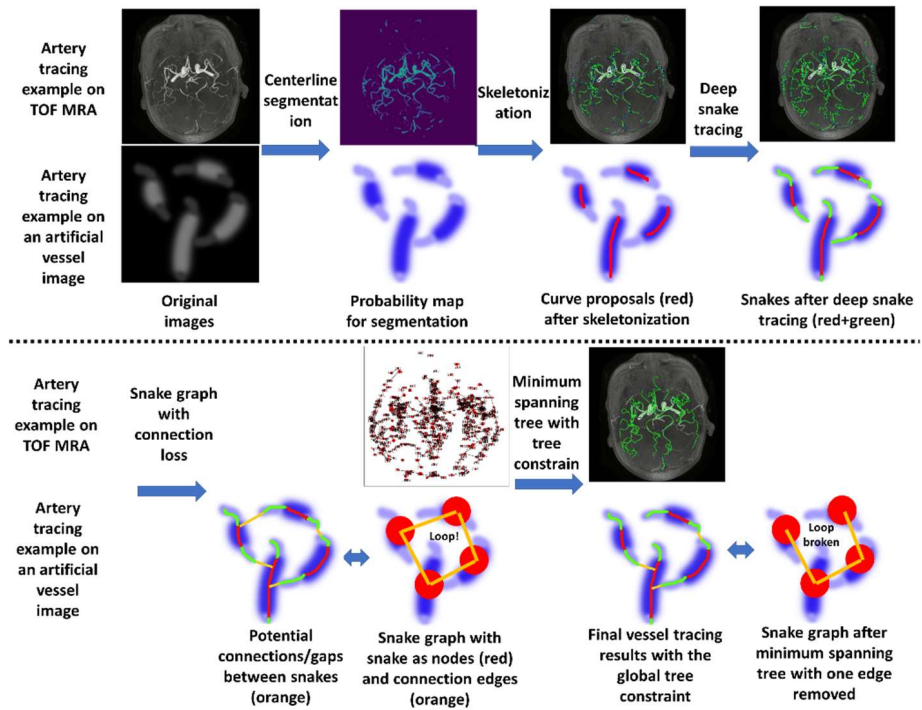

**Fig. 2.** Workflow of DOST. Centerlines for arteries were identified by a centerline segmentation CNN and skeletonized into pieces of 1-voxel thin vessel curves, which were used as the initial traces for deep snake tracing. CNN predicted the stretching directions and radii for both ends of the snake while trace smoothness and fitness to image intensities were maintained. Then a graph model was used to select and construct a topologically correct vascular tree.

The tracing target for each scan in our task is a constructed snake list $T = \{t_i, i = 1, 2, \ldots, N\}$ with $N$ traces. Ground truth labels are annotated with hat symbols, for example, $\hat{T}$.

Each snake is a list of points with 3D position $\boldsymbol{p}_{i,j} = (x_{i,j}, y_{i,j}, z_{i,j})$ along with the corresponding radius of lumen area $r_{i,j}$.

$$t_i = (P_i, R_i) = (\{\boldsymbol{p}_{i,j}, j = 1, 2, \ldots, n\}, \{r_{i,j}, j = 1, 2, \ldots, n\}) \quad (1)$$

## 2.2    Curve proposal from centerline segmentation

Instead of seed points [12,13], DOST initializes from *vessel curves* predicted from the centerline segmentation, which better utilizes vascular structures from segmentation-based tracking methods to avoid initial stretching errors. A 3D patch-based encoder-decoder centerline segmentation network (2 blocks in encoder/decoder, each with 2 3D convolutional layers+RELU followed by 3D max pooling/up sampling layers) was used for segmentation on the vascular images. To separate the nearby vessels, the centerline distance transform [11] was used to map the vascular regions with continuous values according to the distance to the centerline. Each voxel at $\boldsymbol{p} = [\boldsymbol{x}, \boldsymbol{y}, \boldsymbol{z}]$ within the vascular labels was transformed into a labeled distance map $d$, where centerlines voxels had highest values.

$$d[x, y, z] = \max_i \max_j \left( \frac{\max{(0, r_{i,j} - \|\boldsymbol{p}_{i,j} - \boldsymbol{p}\|)}}{r_{i,j}} \right) \quad (2)$$

Considering the majority of $d$ is zeros. The L2 loss for training the segmentation network is masked by the non-zero regions in $d$. $\mathbf{1}(\cdot)$ is an indication function.

$$Loss_{seg} = \|\mathbf{1}(d[x, y, z] \neq 0)(d[x, y, z] - \hat{d}[x, y, z])\|_2 \quad (3)$$

For each dataset, an individual segmentation network was trained with initial weights from a meta-segmentation model [17] trained with our previous MRA data [6,10,18] (no overlap with datasets used in this study) so that the adaptation was efficient.

Threshold after segmentation was chosen (from the validation set) to make binary predictions. Zhang's skeletonization algorithm [19] was applied to identify initial curves (grouped if voxels were 26-connected) for tracing.

## 2.3    Deep snake tracing

An initial curve $\boldsymbol{c}(s)$ was represented as a parametric open curve model $(x(s), y(s), z(s)), s \in [0,1]$ The snake energy used in DOST $E_{DOST}$ was a combination of the internal energy $E_{int}$ and the external energy of $E_{ext}$

$$E_{DOST} = \int_0^1 E_{int}(\boldsymbol{c}(s)) + E_{ext}(\boldsymbol{c}(s)) \, ds \quad (4)$$

$$E_{int}(\boldsymbol{c}(s)) = \alpha(s)|\boldsymbol{c}_s(s)|^2 + \beta(s)|\boldsymbol{c}_{ss}(s)|^2 \quad (5)$$

"Elasticity" $\alpha(s)$ and "stiffness" $\beta(s)$ of the snake were set to be zero at $s = 0 \; or \; 1$ to allow snake stretching. $c_s$ and $c_{ss}$ indicate first and second order derivatives.

$$E_{ext}(\boldsymbol{c}(s)) = -I(x(s), y(s), z(s)) + E_{str}(\boldsymbol{c}(s)) \quad (6)$$

$I$ is the image intensity. $E_{str}$ is the energy for stretching directions at both ends of the snake. $E_{str}(\boldsymbol{c}(s)) = 0$ for points other than both ends.

Different from the OCS, the stretching directions at the end of the snake were predicted from a deep neural network following the settings from the CNN tracker[13]. The network structure has 6 blocks followed by a convolution layer with $D + 1$ dimensions ($D = 500$), each block was a 3D convolutional layer+batch normalization+RELU. A 3D image patch $P(\boldsymbol{c}(s))$ with the size of $19^3$ (large enough to cover the vessel diameter) was extracted centered at $\boldsymbol{c}(s), s = 0, 1$ for prediction of the 1-dimensional radius $r(P(\boldsymbol{c}(s)))$ and $D$-dimensional stretching direction $\boldsymbol{v}_{m=1,2,...,D}(P(\boldsymbol{c}(s)))$ by the network. The predicted stretching magnitudes $\{k_m\}$ has $D$ dimensions indicating evenly distributed 3D unit directions $\boldsymbol{v}_m$. The training targets for each patch were generated from semi-automatically traced arteries using iCafe[20].

$$\nabla E_{str}\big(c(s)\big) = - \begin{cases} \max_m(sign(-\boldsymbol{c}_s(s) \cdot \boldsymbol{v}_m) \cdot k_m), \text{s} = 0 \\ \max_m(sign(\boldsymbol{c}_s(s) \cdot \boldsymbol{v}_m) \cdot k_m), \text{s} = 1 \\ 0, 0 < s < 1 \end{cases} \tag{7}$$

The directions of $-\boldsymbol{c}_s(s)|s = 0$ and $\boldsymbol{c}_s(s)|s = 1$ pointed outward from the curve, indicating the correct stretching directions for $\boldsymbol{v}_m$ to stretch the snake.

Snakes were traced independently. In each iteration for a snake tracing, the snake stretched at both ends with the step size of $\gamma = 0.5r(P(\boldsymbol{c}(s)))$ with minimized snake energy, then resampled evenly for the next iteration.

Snake stretching was terminated either when a snake end point reached another traced snake or the predicted direction $\boldsymbol{v}_m$ can no longer predict a confident direction indicated by the normalized entropy [13].

$$H = \frac{\sum_m -k_m \cdot log_2(k_m)}{log_2(D)} \tag{8}$$

### 2.4 Global tree construction

Deep snake tracing can reliably trace individual arteries. However, traces were not connected with each other through merging or branching to form a topologically meaningful vascular tree. In addition, the global constraint on tree structure (no loop) can be used to fix connection errors in tracing. On rare occasions when loops naturally exist, such as in individuals with collateral arteries or with a complete circle of Willis, a manual extra step of loop reconnection can be added.

The vascular tree was constructed using an undirected *snake graph*, in which vertices indicated snakes, and edges between vertices indicated the connection loss $Loss_{con}(i, j)$ for the snake pair $i, j$ (through merging or branching).

The connection loss $Loss_{con}(i, j)$ was based on intensities of point list $t_{i,j}$ from two snakes $\boldsymbol{c}_i \cup \boldsymbol{c}_j$ and their gap $g_i$ (minimum distance from one point of the snake to any point on the other snake). Foreground intensities for the snake pair were estimated with normal distributions $N_{i_f, \delta_f}$. $i_f, \delta_f$ were mean and standard deviation of intensity along $t_{i,j}$. Background intensities for the snake pair were sampled from points at twice the radius around $t_{i,j}$ with normal distributions $N_{i_b, \delta_b}$. $i_b, \delta_b$ were mean and standard deviation of background intensities. Mean intensity along $g_i$ was $i_g$.

$$Loss_{con}(i, j) = \frac{N_{i_f, \delta_f}(i_g)}{N_{i_f, \delta_f}(i_g) + N_{i_b, \delta_b}(i_g)} \tag{9}$$

When $Loss_{con}(i,j)$ was below the threshold of 0.05 or the gap was above the maximum distance for connection consideration (10mm), edges were removed from the graph. Parameters were empirically chosen for the best vascular tree construction.

Kruskal's algorithm [21] was used for minimum spanning tree (MST) construction from the snake graph with connection losses. Edges in MST were used to fill the gaps for snakes to construct a whole vascular tree.

## 3 Experimental settings and results

### 3.1 Datasets

We used the BRAVE dataset [5,22] with 167 TOF MRA from elderly hypertensive subjects to evaluate DOST comparing with state-of-the-art methods. To evaluate the robustness of DOST on more vascular beds and imaging modalities, we used 1) the Rotterdam Coronary Artery Challenge (CAT08): eight CTA scans for coronary artery tracing [23], and 2) Harborview dataset: clinical scans with MR vessel wall imaging using T1 SPACE (black blood) from 15 patients with history of stroke. Detailed dataset properties are described in **Table 3** (supplementary material). Ethics approval was waived due to the retrospective study design.

Ground truth for CAT08 was provided. For other datasets, arteries were automatically generated by the OCS and manually corrected in an analysis tool of iCafe [20,24].

### 3.2 Evaluation metrics

We evaluated the tracing accuracy following the metrics used in the CAT08 Challenge [23], including Average inside (AI) and Overlap (OV). However, AI and OV mainly evaluate centerline overlap for a single artery and cannot reflect the multi-vessel connection accuracy. For example, a vessel matched by two traces (**Fig. 1** (b)) has 100% OV but gets no penalties on additional broken predictions.

Thus following multiple object tracking tasks [25], we adopted three 3D multi-vessel connection accuracy metrics: ID switch (IDS), multiple object tracking accuracy (MOTA) and IDF1 metrics to evaluate connection errors in vessel tracing. After $\{t_i\}$ were matched ($min_{i,j}(||\boldsymbol{p}_{i,j} - \hat{\boldsymbol{p}}_{i,j}|| < \hat{r}_{i,j})$) with $\{\hat{t}_i\}$, for each $\hat{t}_i$, ID switch $IDS_i$ was defined as the additional count of sources from the predicted $\{t_i\}$. $TP$ was the number of $\{\hat{t}_i\}$ having matching $t_i$, $FN$ was the number of $\{\hat{t}_i\}$ having no matching $t_i$, and $FP$ was the number of $\{t_i\}$ having no matching $\hat{t}_i$. $T$ was the number of points in $\{\hat{t}_i\}$. $MOTA$ penalizes $FP$, $FN$ as well as IDS, leading to the best possible value of 1. $IDF1$ is the F1 score for artery matching ranging from 0 to 1, with lower score either by larger $FP$ or $FN$.

$$IDS = \sum_i IDS_i \qquad (10)$$

$$MOTA = 1 - \frac{FN+FP+IDS}{T} \qquad (11)$$

$$IDF1 = \frac{2 \cdot TP}{2 \cdot TP + FP + FN} \qquad (12)$$

### 3.3 Evaluation on BRAVE

Comprehensive quantitative comparisons with DOST were made with traditional and deep learning methods. For segmentation-based approaches, Frangi vesselness filter [26], U-Net [27] and Deep Distance Transform (DDT) [11] were selected. For tracking-based approaches, open curve snake (OCS) [16], CNN-tracker [13] and Discriminative Coronary Artery Tracking (DCAT)[14] were selected. Results were shown in **Table 1**.

All experiments were implemented in PyTorch 1.4.0 and TensorFlow 1.15.0 on a workstation with an Intel Xeon E5-1650 v4 CPU and a NVIDIA Titan V GPU.

**Table 1.** Quantitative comparison results for intracranial artery tracing

| Tracing approach | Model name | OV ↑ | AI ↓ | MOTA ↑ | IDF1 ↑ | IDS ↓ |
|---|---|---|---|---|---|---|
| Traditional segmentation | Frangi | 0.617 | 0.956 | 0.238 | 0.621 | 343.9 |
| Deep learning segmentation | U-Net | 0.662 | 0.724 | 0.300 | 0.696 | 398.3 |
| Deep learning segmentation | DDT | 0.683 | 0.703 | 0.281 | 0.712 | 423.0 |
| Traditional tracking | OCS* | 0.672 | **0.356** | **0.372** | 0.694 | **74.8** |
| Deep learning tracking | CNN tracker | 0.562 | 0.860 | -0.312 | 0.595 | 108.5 |
| Deep learning tracking | DCAT | 0.564 | 0.943 | -0.241 | 0.601 | 137.8 |
| Hybrid | DOST (Our) | **0.732** | 0.592 | 0.318 | **0.731** | 104.1 |

\* Ground truth was modified manually based on OCS results

DOST showed higher performance than most methods. Note that the ground truth was generated based on OCS, which was a natural bias. An example of artery tracing on a MRA data is shown in **Fig. 3**.

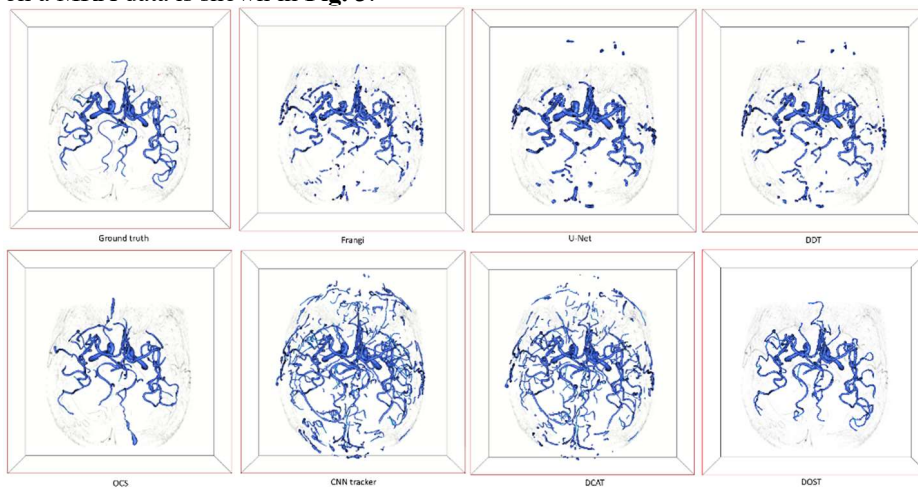

**Fig. 3.** An example of artery tracing for comparison. DOST did not have the problem in segmentation-based method for broken arteries, and it considered global tree structures in tracing, thus avoiding loops or many noise branches.

### 3.4 Ablation study

We tested DOST performance on the BRAVE dataset without key modules of deep snake tracing, and global tree construction. Results are shown in **Table 2**. Deep snake tracing improved overall performance, and global tree constraint mainly improved multi-vessel connection accuracy but with minor drop of OV and AI.

**Table 2.** Ablation study of DOST modules evaluated on BRAVE dataset.

| Modules | OV ↑ | AI ↓ | MOTA ↑ | IDF1 ↑ | IDS ↓ |
|---|---|---|---|---|---|
| Initial curve | 0.655 | 0.781 | 0.274 | 0.702 | 467.0 |
| Initial curve+deep snake tracing | **0.719** | **0.651** | 0.292 | 0.732 | 323.8 |
| DOST (Initial curve+deep snake tracing+global tree) | 0.709 | 0.655 | **0.339** | **0.738** | **120.0** |

### 3.5 Adaptability of DOST on other datasets

On the CAT08 dataset, DOST achieved OV of 99.6% and AI of 0.22, which were comparable with CNN tracker (OV: 98.7% and AI: 0.18). Tracing results are shown in **Fig. 4** (supplementary material).

On the Harborview dataset, DOST achieved OV of 91.7%, AI of 0.69, MOTA of 0.843, IDF1 of 0.925 and IDS of 7.4. For black blood MRI, traditional methods did not work, but DOST was feasible after retraining the centerline segmentation using a small training set (5 cases). An example is shown in **Fig. 5** (supplementary material).

## 4 Discussions and Conclusion

A deep learning based open curve snake model (DOST) was developed and evaluated. DOST combines deep learning-based direction prediction/radius estimation and the classic parametric curve modeling. It allows data driven machine learning knowledge to complement human prior knowledge on the structure of vessels (smoothness and stretchiness) and the topology of the vasculature, so that DOST out-performed existing models with either human or machine knowledge. In addition, DOST, as an adaptive hybrid (segmentation and tracking-based) tracing method, is able to identify complete vascular trees from multiple vascular beds and modalities.

DOST is not purely post processing/smoothing on CNN based tracking results. The snake constraint was applied during the trace stretching and the tracing was initiated from segmentation-based curve proposal instead of seed points.

The main limitation of DOST is its requirement of supervised training to get the initial curve proposals and predict stretching directions. The training labels requires detailed semi-automated artery tracings[20]. However, we have demonstrated DOST works even with 6 cases in CAT08 and 5 cases in Harborview dataset.

With the combination of artery labeling [28], vessel wall segmentation [29], feature quantification [20] and visualization [30], a complete workflow for 3D vascular map construction and analysis can be automated, which will greatly benefit quantitative vascular research and has much potential on clinical diagnosis on vascular diseases.

## 5    Acknowledgement

This work was supported by National Institute of Health under grant R01-NS092207. We are grateful for the collaborators who provided the datasets for this study, including the BRAVE investigators, Harborview Medical Center, and the public data from Erasmus MC, Rotterdam. We gratefully acknowledge the support of NVIDIA Corporation for donating the Titan GPU.

# Supplementary material

**Table 3.** Detailed properties for datasets used in our study.

| | Number of training/validation/teat cases | Modality | Vascular beds | Blood intensity in images | Interpolated in-plane resolution (mm) |
|---|---|---|---|---|---|
| BRAVE | 117/25/25 | TOF MRA | Intracranial | Bright | 0.43 |
| CAT08 | 6/1/1 | CTA | Coronary | Black | 0.36 |
| Harborview | 5/5/5 | MR VWI (T1 SPACE) | Intracranial | Bright | 0.56 |

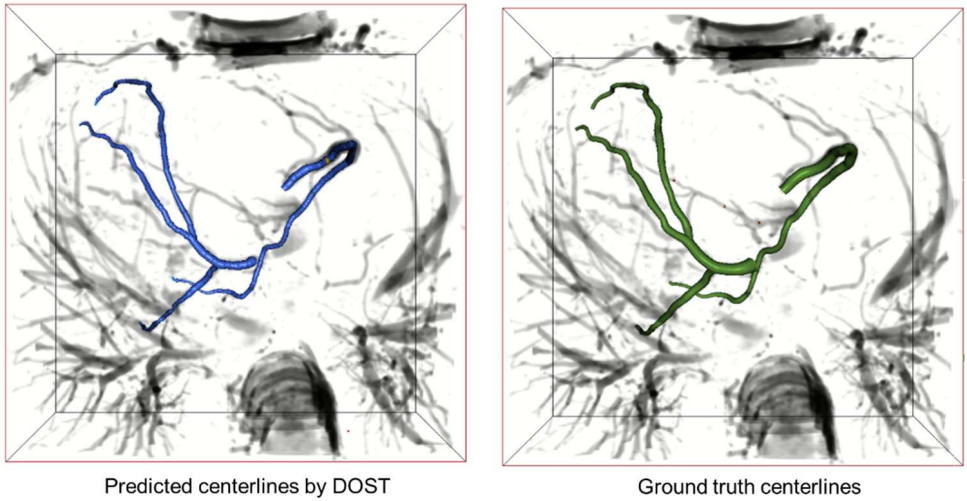

Predicted centerlines by DOST          Ground truth centerlines

**Fig. 4.** Coronary artery tracing results from DOST compared with ground truth.

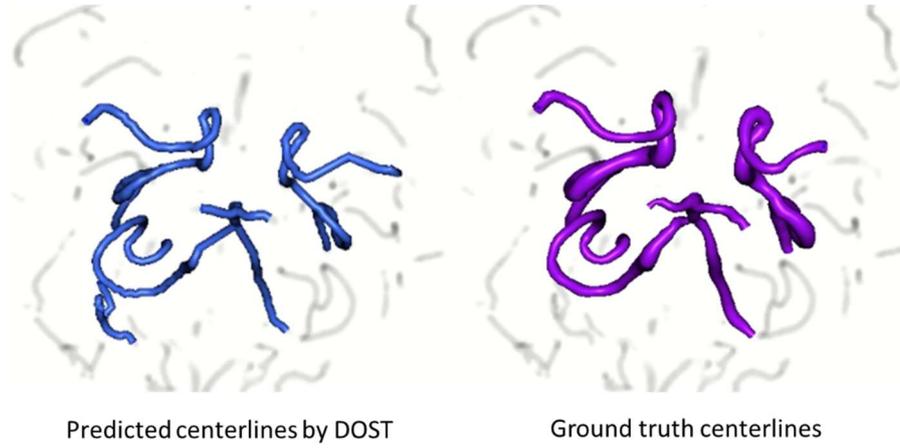

Predicted centerlines by DOST          Ground truth centerlines

**Fig. 5.** Intracranial artery tracing from T1 vessel wall images using DOST compared with ground truth.